\documentclass[aps,prl,superscriptaddress,twocolumn,tightenlines]{revtex4}

\usepackage {graphicx}
\usepackage {amsmath}
\usepackage {amsfonts}
\usepackage {amssymb}
\usepackage {mathrsfs}

\newcommand {\be}{\begin{eqnarray}}
\newcommand {\ee}{\end{eqnarray}}
\newcommand {\rmd} {{\rm d}}

\begin{document}

\title{Ballistic properties of crystalline defects}
\author{V. Cvetkovic}
\email{vladimir@pha.jhu.edu}
\affiliation{Department of Physics and 
Astronomy, The Johns Hopkins University, Baltimore, MD 21218}
\author{Z. Nussinov}
\email{zohar@wuphys.wustl.edu}
\affiliation{Department of Physics. Washington University, MO 63160, USA}
\author{J. Zaanen}
\email{jan@lorentz.leidenuniv.nl}
\affiliation{Instituut Lorentz voor de theoretische natuurkunde, 
Universiteit Leiden, P.O. Box 9506, NL-2300 RA Leiden, 
The Netherlands}

\date{\today}
\begin{abstract}
We introduce a field theoretic formalism 
enabling the direct study of dislocation and interstitial dynamics. 
Explicit expressions for the energies of such defects are given.
We provide links to earlier numerical, discrete elastic,
time dependent Ginzburg Landau, and other approaches 
sought by numerous authors for the problem of 
defect dynamics. The formalism introduced in 
this article may be 
extended to address many other systems. Apart from their 
heavily studied role in dislocation mediated crystal melting,
defect dynamics recently gained much interest
due to their viable role in various electronic and other
systems of current interest. These systems include,
yet are not limited to, stripe phases of Quantum Hall liquids,
theories addressing the melting of stripe phases 
in various doped oxides 
(including the high temperature cuprate superconductors,
the nickelates, and manganates), defects in supersolids,  
as well as colloidal crystals.
\end{abstract}

\pacs{ ...... }

\maketitle

{\it Introduction} ---  The formulation of a 
field theoretical framework for analyzing 
{\em defect dynamics} in two dimensional crystals is the central goal
of this work. Defects such as
interstitials/vacancies, dislocations and disclinations 
\cite{Friedel, Nabarro, Kleinert}
perturb the ideal perfectly periodic crystal.
%In some solids, such as shape memory alloys \cite{shape_memory}, 
%reducing the appearance of defects is a matter of pertinence.
To date, numerous works extensively studied 
the dynamics of these defects, e.g., \cite{Hatano,CB, KHEG, ASV, Sharma},
and some further considered how 
these may, in turn, impede diffusion, e.g. \cite{deem}.
Although much progress has been made in analyzing defect dynamics,
a practical and general field theoretical approach that is free from the direct
analysis of the equations
of motion 
is lacking. Such a general framework will enable
access to numerous physical systems
of empirical interest.
In the current publication,
we report on precisely such an approach. For simplicity,
we focus on dislocations
and interstitials in
two dimensional solids free from non-harmonic
terms and dissipation.
The extension to three
dimensions is straightforward. The inclusion of
disclinations is much more involved. 
In the two dimensional
case, the defects are 
point-like.  As these excitations 
are well localized, we can wonder if 
they exhibit the behavior of ballistic particles.
For instance, {\em what is their energy when in motion?}
The answer to this question
is found in Eqs.\ (\ref {Eint}, \ref {Edisl}), our main result showing
how both interstitials and dislocations run
in a 'sound barrier' set by the phonon velocities
in a way that is however
different from  a simple Lorentz invariance.

The accumulation of the defects whose dynamics
we study here is commonly thought
to melt the solid phase. Early on, Onsager \cite{Onsager} theorized that 
the $\lambda$ transition in liquid $^{4}$He 
is driven by an unbinding 
of vortex lines. Shockley \cite{shock} suggested 
that the melting of solids, in general, is driven by the proliferation 
of dislocations. These microscopic 
considerations were couched in
a field theoretical formulation by Kleinert \cite{Kleinert}.
Empirical studies of these notions have been 
carried out, e.g., \cite{burakovsky}. 
A scenario extending 
the two dimensional melting 
of  \cite{KT} which 
further suggested an intermediate 
phase driven by the proliferation of
defects was advanced in \cite{NHY, nelson}. 
The glass transition has also been suggested
to be related to defect dynamics \cite{glass, tarjus, st}.
In glasses, an extensive configurational
entropy of these defects \cite{glass} 
(from which ensuing restrictive slow 
dynamics might follow \cite{wolynes})  
may be sparked. Investigations of 
Frank Kasper phases \cite{FK1} of defect lines and of
related systems \cite{FK2, FK3, FK4, FK5, FK6, FK7, FK8}
have flourished. Recent
years saw the extension of these ideas
to various strongly correlated electronic systems
in which the electronic constituents favored
the formation of an ideal crystal like
stripe pattern. Such 
patterns were detected in the high temperature
superconductors and other oxides \cite{tranquada}
and in quantum Hall systems \cite{qhe}.
Other systems harboring
ideal stripe order 
include ferrofluids, diblock copolymers
\cite{andelman}, and colloidal crystals
\cite{der, reich}.  In an interesting development, recent 
experiments on solid Helium-4 \cite{supersolid} suggested supersolid-type features
(a solid analogue of the superfluid phases) \cite{supersolid-theory} and have led to a flurry of activity. Elastic defects \cite{supersolid-theory, Balibar} 
may give rise to the observed phenomena either directly as superfluid
or indirectly \cite{glassHe4}.
 Indeed, experimental results show \cite{Reppy}  that the measured putative supersolid-type feature is acutely sensitive to the quench rate. The results that we report here pertain to defect dynamics in 
 these disparate arenas. Specifically, our results hold for
all classical incarnations or bosonic variants
in the quantum arena of these crystals.
In earlier works we employed the general framework related to the one used here.
This  enabled us to study other facets of defect dynamics earlier and amongst other results,
allowed to also find an elastic analogue of the Higgs effect  \cite{ZMN, CZ, electric, thesiscvetkovic}
and to derive the glide constraint of dislocations and to determine new non-linear
corrections to it \cite{glide, thesiscvetkovic}.

{\it Elastic gauge fields} --- Deformations in a
two dimensional crystal are parameterized by 
a displacement field ${\bf u}$ monitoring
the deviations of each constituent ``atom'' relative
to its location in the ideal crystal.
We begin our analysis within Euclidean
space (with $\tau = i t$ the imaginary time) 
where the Lagrangian density
\be
  {\cal L} = \tfrac 1 2 \partial_\mu u^a C_{\mu \nu a b} 
\partial_\nu u^b + ... \label {L0}
\ee
The Greek indices $\mu, \nu = 0, 1, 2$ with $\mu=0$ (or $\nu) =0$ 
denoting a temporal component while $a,b = 1,2$ correspond to  
spatial directions alone. To account for the dynamics of the 
constituent particles, a kinetic term borne by $C_{0 0 a b} 
\equiv \rho ~ \delta_{ab}$ 
is inserted ($C_{0 i a b} =0$ for $i \neq 0$).
The spatial components of the elasticity tensor $C_{ijab}$
host the usual harmonic strain energies
($V^{(1)}=\tfrac 1 2 \partial_i u^a  C_{ijab} \partial_j u^b$). 
With the kinetic 
($ C_{\tau \tau a b}$) term in tow, the variational 
equations immediately 
provide the classical dynamics.  
In the quantum arena, this kinetic term
is introduced similarly for bosonic matter 
wherein the partition function
\begin{eqnarray}
Z = \int D {\bf u} ~ e^{-\int d^{2}x d \tau {\cal L}},
\end{eqnarray}
We analyze the 
isotropic case, $C_{ijab} = 
\mu \lbrack \delta_{ij} \delta_{ab} + \delta_{ib} 
\delta_{ja} + \frac {2 \nu}{1 - \nu} \delta_{ia} \delta_{jb} \rbrack$ 
\cite {comment0}. 
At short distances, terms containing higher derivatives become 
significant. %, and we have to include them in our analysis in order
%to avoid divergencies.
For the studied isotropic case only two additional terms are
symmetry allowed. %, regardless  of the point group of the crystal.
The second order gradient 
energy density augmenting the first term of
Eq.(\ref{L0}) is 
\begin{eqnarray}
V^{(2)} = 2 \mu \ell^2 (\nabla \omega)^2
+ \frac {\mu}{1 - \nu} \ell'^2 
(\nabla (\nabla \cdot {\bf u}))^2 \label{V2.}
\end{eqnarray} 
The length $\ell$, 
relates to the local rotation 
($\omega \equiv  \tfrac 1 2 \nabla \times {\bf u}$) inhomogeneities. 
At distances comparable to length $\ell'$, the compression 
gradient becomes significant.
As customary in studying defects elsewhere, we 
dualize the action. Stresses are dual to displacements and  
are given by 
\be
  \sigma_\mu^a &=& i \frac {\partial {\cal L}}{\partial 
(\partial_\mu u^a)} = i C_{\mu \nu a b} \partial_\nu u^b, \nonumber \\
  \tau'_i &=& i \frac {\partial {\cal L}}{\partial (\partial_i 
(\nabla \cdot {\bf u}))} = i \frac {2 \mu}{1 - \nu} \ell'^2 \partial_i 
(\nabla \cdot {\bf u}),\\
  \tau_i &=& i \frac {\partial {\cal L}}{\partial (\partial_i \omega)} 
= i 4 \mu \ell^2 \partial_i \omega. \nonumber
\ee
The temporal components $\sigma_\tau^a$ are the physical momenta, and the 
spatial $\sigma_i^a$ are elastic stresses. The Hamiltonian density, 
${\cal H} = i \sigma_\mu^a \partial_\mu^a + 
i \tau'_i \partial_i (\nabla \cdot {\bf u}) + i \tau_i \partial_i \omega 
+ {\cal L}$, becomes
\be
  {\cal H} = \tfrac 1 2 \sigma_\mu^a (C_{\mu \nu a b})^{-1} 
\sigma_\mu^a + \tfrac 1 2 \frac {1 - \nu}{2 \mu \ell'^2} 
\tau'_i \tau'_i + \tfrac 1 2 \frac 1{4 \mu \ell^2} \tau_i \tau_i.
\ee
A (rotationally related) singularity present in this form is
readily removed by enforcing 
the Ehrenfest constraint 
\begin{eqnarray}
\epsilon_{ab} \sigma_a^b = 0,
\label{Ehrenfest}
\end{eqnarray} 
(with $\epsilon_{12} = - \epsilon_{21} = 1$).
The dual action is recovered from the inverse Legendre transform 
${\cal L}_{dual} = - i \sigma_\mu^a \partial_\mu^a - i \tau'_i \partial_i 
(\nabla \cdot {\bf u}) - i \tau_i \partial_i \omega + {\cal H}$. 
The displacement has smooth and singular parts 
${\bf u} = {\bf u}_{smooth} + {\bf u}_{singular}$ with 
${\bf u}_{singular}$ carrying the defect contribution. 
Upon integration over ${\bf u}_{smooth}$, we are led 
to the constraint
\be
  0 = \partial_\mu \left \lbrack \sigma_\mu^a - \delta_{\mu a} 
\partial_i \tau'_i - \tfrac 1 2 \epsilon_{\mu a} \partial_i \tau_i 
\right \rbrack.
\label{fc}
\ee
Eq.(\ref{fc}) is none other than an expression of
a generalized (space-time) conservation law
(in the simplest equilibrium setting, the spatial portion
of Eq.(\ref{fc}) dictates a divergence-less 
stress $\sigma$ which implies a net zero force acting on each volume 
element).  Eq.(\ref{fc}) enables us to express the 
brackets as a curl of stress gauge fields,
\be
  \sigma_\mu^a = \epsilon_{\mu \lambda \rho} \partial_\lambda B_\rho^a 
+ \delta_\mu^a \partial_i \tau'_i + \tfrac 1 2 \epsilon_{\mu a} 
\partial_i \tau_i.
\label{curlB}
\ee
From Eq.(\ref{curlB}), it is evident that $B$ is defined 
only up to a gauge transformation, ($B^{a}_{\mu} \to 
B^{a}_{\mu} + \partial_{\mu}f^{a}$ with arbitrary $f^{a}$ leave 
$\sigma_{\mu}^{a}$ unchanged).
In what follows, we compute in the Coulomb gauge
$\partial_i B_i^a = 0$ where the calculations 
simplify considerably.
After the partial integration, the singular 
components become minimally coupled to
the dislocation currents to the gauge fields,
leading to $[- i B_\mu^a J_\mu^a]$,
wherein the dislocation currents 
\begin{eqnarray}
J_\mu^a = \epsilon_{\mu \rho \lambda} \partial_\mu \partial_\lambda u^a.
\end{eqnarray}
Following a Fourier transformation, we introduce a new basis. For fields 
of spatial momentum ${\bf q}$  (and frequency $\omega$), 
we define two planar basis vectors- a 
longitudinal ${\bf \tilde{e}}^L = \frac {\bf q}{|{\bf q}|} = (\cos \phi, 
\sin \phi)$ 
and a 
transversal 
${\bf\tilde{e}}^T = \frac {{\bf q} \times}{|{\bf q}|} = (-\sin \phi, \cos 
\phi)$.
Expressed in the new basis, the gauge fields are $B_\tau^a = 
i \tilde{e}_a^E B_\tau^E$ (double indices are summed over 
with $E= L,T$)  and 
$B_i^a = - e_i^M \tilde{e}_a^E B_M^E$  (with $M= \tau, T$). 
In this basis,
the Coulomb gauge fix condition becomes $B_L^E = 0$. 
Only the 
longitudinal components of the fields $\tau'$ and $\tau$ have
a physical meaning. This allows us to omit 
the transversal components. 
Finally, the Ehrenfest 
constraint (Eq.(\ref{Ehrenfest})) may be employed 
to express $\tau = - B_\tau^L + i \frac{\omega}{q} 
B_T^T$, with $q \equiv \sqrt{|{\bf q}|^{2} + \omega^{2}}$.
Armed with the above constraints, an 
integration over the non-propagating
fields, $\tau'$ and  
$B_\tau^E$, leads to 
\begin {widetext}
\be
  {\cal L}_B &=& \tfrac 1 2 B_T^{L \dagger} \frac {\frac {\omega^2}
{c_L^2} + q^2 (1 + \ell'^2 q^2)}{\rho (1 + \ell'^2 q^2)} B_T^L + 
\tfrac 1 2 B_T^{T \dagger} \frac {\frac {\omega^2}{c_T^2} 
+ q^2 (1 + \ell^2 q^2)}{\rho (1 + \ell^2 q^2)} B_T^T +
\tfrac 1 2 J_\tau^{T \dagger} \frac {2 \mu}{q^2} 
\frac {1 + \nu + 2 \ell'^2 q^2}{1 + \ell'^2 q^2} J_\tau^T \nonumber \\
  &&+ \tfrac 1 2 
J_\tau^{L \dagger} \frac {4 \mu \ell^2}{1 + \ell^2 q^2} J_\tau^L 
- i B_T^{L \dagger} \frac {(J_T^L - \nu J_L^T) 
+ \ell'^2 q^2 (J_T^L - J_L^T)}{1 
+ \ell'^2 q^2} 
- i B_T^{T \dagger} \frac {(J_T^T - J_L^L) + \ell^2 q^2 
(J_L^L + J_T^T)}{1 + \ell^2 q^2}. \label {LB}
\ee
\end {widetext}
Eq.(\ref{LB}) constitutes a central result of our work. 
It enables a direct evaluation of energy densities
from microscopic details. We will shortly illustrate
how the energy densities for different defects
(both static and dynamic) 
easily follow from this very general expression.
Let us first comment on the various terms and associated
symmetries. The first two  terms in Eq.(\ref{LB})
contain propagators for the longitudinal 
and transversal ``photon'' (phonon) respectively. The third and 
fourth terms account for the interaction amongst 
static dislocation charges.  Finally, the
last two terms embody the coupling between the elastic photons and 
dynamical dislocation currents. As detailed
elsewhere \cite{glide}, the various current
combinations adhere to different rotational symmetries-
(i) the compression (or glide) current $[J_L^T - J_T^L]$ vanishes 
for dislocations; 
(ii) the current 
$J_E^E = \partial_\tau ( \nabla \times {\bf u} ) - \nabla \times 
(\partial_\tau {\bf u})$ and (iii) the combination $[J_L^T + J_T^L 
\pm i J_T^T \mp i J_L^L]$ is a shear doublet.
We now evaluate, by the insertion of the corresponding 
current densities ($J$) into Eq.(\ref{LB}), the energies of static and 
dynamic defects. We start with the static case. Associated with 
a static dislocation of Burgers vector ${\bf b}$ is a
current $J^{a; disl.}_\tau = b^a \delta ({\bf r})$. Henceforth, 
we assume that ${\bf b} = b {\bf e}_x$. An interstitial can be viewed 
as a dislocation dipole -- a difference of two 
dislocations carrying Burgers vectors of the size of lattice spacing 
($a$), misplaced by a lattice spacing in the perpendicular direction. 
The current of an interstitial $J^{a; int.}_\tau =
a^2 \epsilon_{ab} 
\partial_b \delta ({\bf r})$. In Fourier space, 
\begin{eqnarray}
\tilde J^{x; disl.}_\tau(q) = b,
~ ~ ~ \tilde J^{T; int.}_\tau(q) = q a^2.
\end{eqnarray}
When these external currents are inserted into the action of 
Eq.(\ref {LB}), we find the static energy of a single 
dislocation/interstitial. The range of integration
over the momenta $q$ in all pertinent integrals
is $1/L< |q|< 1/\Delta$ with $L$ the linear
dimension of the system size and 
$\Delta$ the defect core size. We facilitate our calculation by setting
$\Delta = a$.
For a single static interstitial (where an infrared cut-off is not 
imperative), the energy derived from Eq.(\ref{LB}), 
the `action per unit time', is 
\be
  E_{static}^{int.} = 2 \pi \mu a^2 \left \lbrack 1 
- \tfrac {1 - \nu}2 \frac {a^2}{\ell'^2} 
\ln (1 + \frac {\ell'^2}{a^2}) \right \rbrack. \label {Lpotint}
\ee
For static dislocations, the energy depends on higher order couplings 
and must be regularized by an IR cut-off,
\be
  E_{static}^{disl.} = \pi \mu b^2 \left \lbrack \ln \left 
( \frac {L^2}{a^2} \frac {1 + \frac {\ell^2}{a^2}}{1 
+ \frac {\ell^2}{L^2}} \right ) + \tfrac {1 - \nu}2 
\ln \frac {a^2 + \ell'^2}{L^2 + \ell'^2} \right \rbrack. 
\label {Lpotdisl}
\ee
The dislocation unbinding energy grows monotonically with the increase of 
the rotational gradient moduli $\ell$ (Eq.(\ref{V2.})) 
yet is bounded from above. By contrast, 
the disclination energy $E_{static}^{disl.}$  grows with with the square 
of the same coupling. (The latter may be verified by a simple 
rotationally symmetric ansatz.) Thus, high values of the 
rotational stiffness $\ell$ suggest a new, 
intermediate phase- in accordance with the 
known results \cite {NHY, ZMN}.

We next employ Eq.(\ref{LB}) to determine the action of
moving defects. For a uniformly moving defect, 
the delta function constraint of the temporal component of the current 
is boosted to $\delta ({\bf x} - 
{\bf v} \tau) = \delta (x - v \tau) \delta (y)$. For the interstitial 
motion in all directions is possible. By contrast, 
single dislocation can only ``glide'', i.e. move parallel
to their Burgers' vectors \cite{Friedel, Nabarro, Kleinert, 
ZMN, glide}. As a consequence, in dealing with 
dislocations we must consider only velocities parallel to 
the Burgers vector. After Fourier 
transformation, the temporal currents for an interstitial and 
a dislocation are, respectively,
\begin{eqnarray}
\tilde J_\tau^{T; int.}(q) = q a^2; ~ \tilde J_\tau^{x;disl.}(q) 
= b \delta (\omega - v q_x)
\delta 
(\omega - v q_x). \nonumber
\end{eqnarray} 
The corresponding spatial components are
\begin{eqnarray} 
\tilde J_x^{T; int.} = i v q a^2 
\delta (\omega - v q_x), ~ ~ 
\tilde J_x^{x;disl.} = v b \delta (\omega - v q_x). \nonumber
\end{eqnarray}
The action per unit time of a kinetic interstitial
exceeds that of a static interstitial by 
\be
  E_{kin}^{int.} \approx \tfrac 1 2 v^2 \rho 
\pi a^2 \nonumber
\\ \times \left \lbrack 1 + \frac {(1 + \nu)^2}{2 (1 + 
\frac {\ell'^2}{a^2})} + (1 + \nu) \frac {a^2}{\ell'^2} 
\ln (1 + \frac {\ell'^2}{a^2}) \right \rbrack. \label {Lkinint}
\ee
The analogous `kinetic'  action density
for a dislocation is
\be
  E_{kin}^{disl.} \approx v^2 \rho b^2 \tfrac \pi 4 \nonumber
\\ \times \left \lbrace \left \lbrack \frac 3{1 + \frac {L^2}{\ell^2}} - 
\frac 3{1 + \frac {a^2}{\ell^2}} + \ln \frac {L^2 + \ell^2}
{a^2 + \ell^2} \right \rbrack + \right . \nonumber \\
  +  \left . \left ( \tfrac {1 - \nu}2 \right )^2 \left 
\lbrack \frac 1{1 + \frac {L^2}{\ell'^2}} - \frac 1
{1 + \frac {a^2}{\ell'^2}} + \ln \frac {L^2 + \ell'^2}
{a^2 + \ell'^2} \right \rbrack \right \rbrace 
\label {Lkindisl}.
\ee

{\it `Covariant' mapping of a defect gas} --- 
With the action densities for both static and dynamic
defects at our disposal, we may now 
map our system onto a Ginsburg-Landau-Wilson 
action of the form employed to describe a gas of
random loops
\be
  {\cal S}_{GLW} =  \int \rmd {\bf x} d \tau
~ \left \lbrack \tfrac 1 2|\partial_\mu \psi|^2
+ \tfrac 1 2m^2 |\psi|^2 
+ \frac{\lambda}{4!} |\psi|^4 \right \rbrack. \label {SGLW}
\ee
With imaginary time as an extra dimension, 
point defects become world lines and the action (\ref {SGLW}) 
is valid as a description of a defect gas. The mapping procedure leading 
to Eq.(\ref {SGLW}), employs a 
representation of the defect gas where, 
upon conversion with some (yet unknown) velocity $c_d$ 
($x_{0} = c_{d} \tau$),
the action of a defect is proportional to its world-line 
length. We will shortly relate $c_{d}$ to the phonon velocity.
A defect, moving with velocity $v$ during a 
time $\Delta \tau$, contributes to the action an amount
\be
  \Delta S = \epsilon_0 ~ \Delta \tau \sqrt 
{1 + \frac {v^2}{c_d^2}} \approx ~ \Delta \tau [\epsilon_0 + \tfrac 
1 2 \epsilon_0 \frac {v^2}{c_d^2}],
\ee
with $\epsilon_{0}$ the defect core energy. 
Contrasting this with Eqs.(\ref{Lpotdisl}, \ref{Lkindisl}),
we find (for $\ell, \ell' \lesssim a \ll L$) that
\be
  (c_{d; disl.})^2 = \frac {\mu}\rho \frac {1 + \nu}{1 + \left 
( \frac {1 - \nu}2 \right )^2}. \label {cd}
\ee
Knowing the velocity (Eq.(\ref {cd})), 
the action for the gas of dislocations 
has the derivative term precisely set by
the first term of Eq.(\ref{SGLW}),
\begin{eqnarray}
|\partial_{\mu} \psi|^{2} =  
\frac 1{2 c_d^2} |\partial_\tau \psi|^2 + |\nabla \psi|^2. 
\label {partialpsi}
\end{eqnarray} 
%Similar expressions appear for single interstitials.
The factor of two 
in the denominator appears for dislocation as a consequence of the glide 
principle-  motion in a direction transverse
to the burgers vector (climb) is not allowed
for dislocations. Constrained defect motion does not lead to
an effective dimensionality reduction of the defect gas.
Instead, a factor of two for velocity in Eq.(\ref {partialpsi}) 
arises \cite {ZMN}, yet the physical propagators 
exhibit poles corresponding to the physical 
velocity $c_d$.

Similar expressions may be derived within this framework for single
interstitials. The differences are that the factor of two lacks from equivalent
of Eq.\ \eqref {partialpsi} due to the fact that tinterstitials are not constrained
by glide, and that the interstitial velocity takes form
\be
  (c_{d; int.})^2 = \frac \mu \rho \frac {2 (1 + \nu) }{1 + (2 + \nu)^2}.
\ee

{\it Real time ballistics} --- Upon extending our results to
real time, we find that the energy diverges as the velocity 
approaches the velocity of `signal carriers' -- 
phonons, $v \to c_T$ 
(the transversal phonon velocity is smaller and it bounds the speed 
of a defect). As the energy has a more complicated form than in the 
imaginary time formulation, we expose only $\ell = \ell' = 0$ 
limit where we can give exact analytic expressions. 
With the previously defined defect currents, now in real time, 
two gauge fields (phonons) will have the solutions
\be
  B_T^L = \frac \rho{q^2 - \frac {\omega^2}{c_L^2}} 
(J_T^L - \nu J_L^T), ~
  B_T^T = \frac \rho{q^2 - \frac {\omega^2}{c_T^2}} 
(J_T^T - J_L^L). \nonumber
\ee
When these fields are inserted in Eq.(\ref{LB}) and an 
integration over Fourier 
space is performed, we obtain the total 
(including the static part) 
energy of defects
\begin {widetext}
\be
  E^{int.} &=& \pi \mu a^2 \left \lbrack 
\frac {(1 + \nu) (4 - 3 \nu)}{2 (1 - \nu)} 
+ \frac {\nu^2}{1 - \nu} \frac 1 {1 - \beta_L^2} - \frac 1
{1 - \beta_T^2} 
+ \frac 1{2 (1 - \beta_T^2)^{3/2}} \right \rbrack, \label {Eint} \\
  E^{disl.} &=& \pi \mu b^2 \ln \frac L a \left \lbrace \frac 
{\nu (2 - \nu)}
{1 - \nu} + \frac {12 \nu (2 - \nu)}{(1 - \nu)^2 \beta_T^2} 
+ \frac {6 - 4 \beta_L^2}{(1 - \nu) \beta_L^2 \sqrt {1 - \beta_L^2}}
- \frac {7 \beta_T^4 - 20 \beta_T^2 + 12}{\beta_T^2 
{(1 - \beta_T^2})^{3/2}} \right \rbrace. \label {Edisl}
\label{Efinal}
\ee
\end {widetext}
Here, $\beta_{T,L} \equiv \frac{v}{c_{T,L}}$.
As the Lagrangian of Eq.(\ref {L0}) is not ``Lorentz invariant'', 
these expressions do not adhere to a single simple covariant 
form ($\frac 1{\sqrt {1 - \beta^2}}$). Nevertheless, we find that
divergences occur when $\beta \to 1$. This divergence of the energy 
was obtained in our derivation by employing the lowest order terms in the elastic
gradient expansion for the Lagrangian density and need not constitute a fundamental limitation on 
interstitial and other defect motion. Two velocities appearing in the problem 
complicate the energy further -- the low velocity approximation 
yields ballistics ($c_d$) which resemble properties from both sectors. 
On the other side, ``causality'' is driven exclusively by the 
transverse phonon velocity 
as the energy of a dislocation diverges already at that velocity.
For small defect velocities, the real time dynamics is not radically 
different from that
in imaginary time. At high velocities, retardation effects 
(triggered by finite phonon velocities) come to the fore 
and the (real time) energy vividly displays 
the role of ``causality'' in the elastic medium.
Eq.(\ref{Efinal}) illustrates the power of our formalism.
With slight variations and different defect current densities, 
Eq.(\ref{LB}) allows the determination of numerous static and 
dynamic dislocation/interstitial configurations.

{\it Conclusion} --- In this work, we illustrate that the old
problem of defect dynamics which is quite cumbersome within the common 
strain formalism becomes far easier within
the framework introduced here. Much unlike the strain formalism, 
where higher order couplings are necessary to close a set of 
differential equations, 
in the formalism studied in the current article, we
obtain expressions which are exact and fairly trivially. 
To illustrate the power of our formulation,
we compute the
energies of moving dislocations
and interstitials (Eq.(\ref{Efinal})).
With slight variations and different realizations of
defect current densities, Eq.(\ref{LB})
allows the analysis of numerous static and 
dynamic dislocation/interstitial configurations.
As the problem of elastic defect dynamics
has myriad manifestations -- crystals, glasses, 
liquid crystals, several strongly correlated electronic 
systems, and many others -- our study affords a fresh
perspective on the defect dynamics within these
systems.

We conclude with two brief remarks--- (i) The dynamics found in imaginary 
time has a crucial consequence on various properties 
of a special nematic phase of a crystal wherein only dislocations
proliferate \cite{ZMN}. Here, dynamic dislocations alter physical 
measurable quantities such as 
elastic response as well as electric 
response functions \cite{electric}. 
(ii) Within this work, only dislocation defect currents were treated. 
To accommodate disclination defects, we need to define the fields
$B$ of Eq.(\ref {curlB}) 
such that $B_\mu^a = \partial_a A_\mu^L + \epsilon_{ab} 
\partial_b A_\mu^T$. The ``double curl gauge fields'' $A_\mu^{L,T}$
are defined up to innocuous 
shifts, $\delta A_\mu^{L, T} = \partial_\mu g^{L, T}$.

{\it Acknowledgments} --- VC wishes to thank 
Daniel Nogradi for useful discussions.

\bibliographystyle{apsrev}

\begin{thebibliography}{10}
\bibitem {comment0} The Poisson ratio in 2D is $\nu = \frac {\kappa - \mu}{\kappa + \nu}$ taking values $-1 \leq \nu \leq 1$.
\end{thebibliography}

\end {document}